\documentclass[%
onecolumn,
superscriptaddress,
 amsmath,amssymb,
 aps,prfluids
]{revtex4-2}

\usepackage{graphicx}
\usepackage{bm}
\usepackage{hyperref}

\usepackage[svgnames]{xcolor}

\definecolor{tab_blue}{HTML}{1F77B4}
\hypersetup{ breaklinks=true, colorlinks=true, linkcolor=tab_blue, filecolor=tab_blue, urlcolor=tab_blue,
 citecolor=tab_blue,
}

\begin{document}

\title{Dynamics of small bubbles in turbulence in non-dilute conditions}

\author{Xander M. de Wit}
\affiliation{Fluids and Flows group, Department of Applied Physics and J. M. Burgers Centre for Fluid Dynamics, Eindhoven University of Technology, P.O. Box 513, 5600 MB Eindhoven, Netherlands}
\author{Hessel J. Adelerhof}
\affiliation{Fluids and Flows group, Department of Applied Physics and J. M. Burgers Centre for Fluid Dynamics, Eindhoven University of Technology, P.O. Box 513, 5600 MB Eindhoven, Netherlands}
\author{André Freitas}
\affiliation{Deptartment of Physics and INFN, University of Rome ``Tor Vergata'', Italy}
\affiliation{LTCI, Télécom Paris, IP Paris, France}
\author{\\Rudie P. J. Kunnen}
\affiliation{Fluids and Flows group, Department of Applied Physics and J. M. Burgers Centre for Fluid Dynamics, Eindhoven University of Technology, P.O. Box 513, 5600 MB Eindhoven, Netherlands}
\author{Herman J. H. Clercx}
\affiliation{Fluids and Flows group, Department of Applied Physics and J. M. Burgers Centre for Fluid Dynamics, Eindhoven University of Technology, P.O. Box 513, 5600 MB Eindhoven, Netherlands}
\author{Federico Toschi}
\email{f.toschi@tue.nl}
\affiliation{Fluids and Flows group, Department of Applied Physics and J. M. Burgers Centre for Fluid Dynamics, Eindhoven University of Technology, P.O. Box 513, 5600 MB Eindhoven, Netherlands}
\affiliation{CNR-IAC, I-00185 Rome, Italy}

\date{\today}

\begin{abstract}
Turbulent flows laden with small bubbles are ubiquitous in many natural and industrial environments. From the point of view of numerical modeling, to be able to handle a very large number of small bubbles in direct numerical simulations, one traditionally relies on the one-way coupling paradigm. There, bubbles are passively advected and are non-interacting, implicitly assuming dilute conditions. Here, we study bubbles that are four-way coupled, where both the feedback on the fluid and excluded-volume interactions between bubbles are taken into account. We find that, while the back-reaction from the bubble phase onto the fluid phase remains energetically small under most circumstances, the excluded-volume interactions between bubbles can have a significant influence on the Lagrangian statistics of the bubble dynamics. We show that as the volume fraction of bubbles increases, the preferential concentration of bubbles in filamentary high-vorticity regions decreases as these strong vortical structures get filled up; this happens at a volume fraction of around one percent for $\textrm{Re}_\lambda=\mathcal{O}(10^2)$. We furthermore study the influence on the Lagrangian velocity structure function as well as pair dispersion, and find that, while the mean dispersive behavior remains close to that obtained from one-way coupling simulations, some evident signatures of bubble collisions can be retrieved from the structure functions and the distribution of the dispersion, even at very small volume fractions. This work not only teaches us about the circumstances under which four-way coupling becomes important, but also opens up new directions towards probing and ultimately manipulating coherent vortical structures in small-scale turbulence using bubbles.

\end{abstract}

\maketitle

\section{Introduction} \label{sec:intro} 
Bubbles immersed in turbulent flows impart complex, multiscale dynamics that play a central role across many natural and engineered systems, from mixing \cite{mathai2020bubbly,Cardoso2024}, to boiling processes \cite{dhir1998boiling,benam2021review,ding2025review}, to specially tailored bubble curtains \cite{wursig2000development,bacot2022bubble} and bubbly drag reduction \cite{van2007bubbly,park2018bubbly,bullee2020bubbly}. In this large variety of bubbly flows, one encounters bubbles across a wide range of different sizes, which can all give rise to different dynamical effects.

This work focuses on the dynamics of bubbles that are small with respect to the Kolmogorov scale of the turbulent flow (also referred to as microbubbles), are undeformable and are subject to strong turbulence so that gravity plays a negligible role. Semantically, this coincides with the high Froude number, low Weber number limit. In numerical simulations, such bubbles have traditionally been studied in a one-way coupling paradigm, meaning that only the advection from the fluid phase exerted onto the bubble phase is taken into account. Such one-way coupling simulations are able to reproduce phenomena such as preferential concentration, dispersion and clustering \cite{Maxey1987,Crisanti1992,Balkovsky2001,Mazzitelli2003,Calzavarini2008,Toschi2009,bec2010turbulent}. Intuitively, this approximation is justified if the concentration of bubbles is dilute, such that both the collective effect of the back-reaction of the bubbles onto the fluid as well as interactions between bubbles themselves can be neglected. Once the bubble volume fraction increases, however, this approximation is challenged, and more sophisticated methods that take all such interactions into account are warranted. This can be achieved using methods that fully resolve the geometry of both phases and their boundary conditions, such as the Immersed Boundary Method \cite{Peskin1972,Fadlun2000,Orlandi2006,Seo2011,Verzicco2023}, or using approximate point-based methods such as Refs.~\cite{hoomans1996discrete,vreman2009two}. While fully resolving methods offer increased accuracy, they exacerbate the computational cost, making it intractable to scale up to simulations of millions of small particles or bubbles. Recently, we have introduced an efficient novel point-based method for simulating small particles in turbulence in the so-called four-way coupling regime, taking the two-sided momentum exchange between particle and fluid phase into account, as well as excluded-volume interactions between the particles, at only a very limited additional numerical cost with respect to one-way coupling \cite{DeWit2024}.

\begin{figure}[t!]
\includegraphics[width=0.85\textwidth]{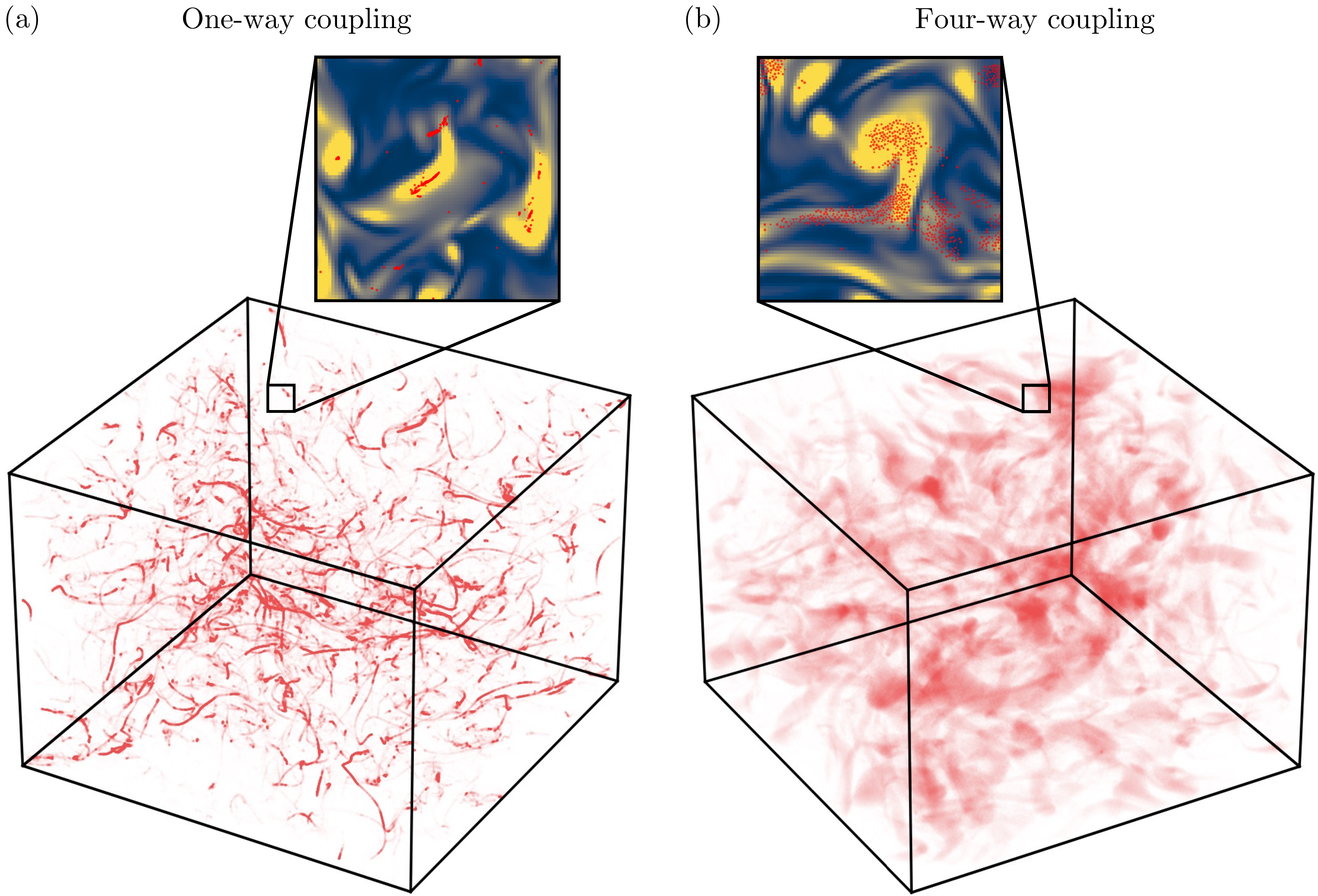}
\caption{\label{fig:4way_example}A snapshot of bubbles in turbulence for a simulation one-way coupled (a) and four-way coupled (b) bubbles of size $D=0.81\eta$ and volume fraction $\alpha=3.5\%$. The inset zooms show a small cross section of the full domain, where the background color indicates the local enstrophy $\Omega^2\equiv|\bm{\nabla} \times \bm{u}|^2$.}
\end{figure}

In this work, we apply this method to simulate undeformable small spherical bubbles in turbulence, see Fig.~\ref{fig:4way_example}. The objective of this work is to study the effect of the interactions included in this four-way coupling paradigm on the dynamics of bubbles in turbulence, spanning volume fractions over five orders of magnitude from a very dilute regime to a very dense regime. We consider various Eulerian and Lagrangian statistical properties that together paint a clear picture of the dynamical behavior of small bubbles in turbulence at the different volume fractions and the influence of four-way coupling thereon. We study under which circumstances, and in which particular observables, effects of four-way coupling emerge, and trace these findings back to the different aspects of the interactions that are accounted for in the four-way coupling method.

\section{Numerical set-up} \label{sec:setup} 
The simulations in this work employ pseudo-spectral Direct Numerical Simulation (DNS) to solve the Navier-Stokes equations for the Eulerian flow in the fluid phase, coupled with Lagrangian interpolation for the bubble phase, solving the Maxey-Riley equations of motion for point particles in turbulence. On top of that, we include a back-reaction of the bubbles onto the fluid phase and excluded-volume interactions between the bubbles, completing the four-way coupling paradigm. The method is discussed and validated in detail in Ref.~\cite{DeWit2024}, here we give a brief description.

The flow field $\bm{u}(\bm{r},t)$ in the fluid phase in Eulerian space $\bm{r}$ and time $t$ is governed by the conservation of momentum and mass as
\begin{subequations}
\begin{gather} \label{eq:NS}
    \frac{\mathrm{D} \bm{u}}{\mathrm{D} t} = -\bm{\nabla} p + \nu \Delta \bm{u} + \bm{f} + \bm{f}_p,\\
    \bm{\nabla}\cdot\bm{u}=0,
\end{gather}
\end{subequations}
where $p$ denotes the pressure and $\nu$ is the kinematic viscosity. The flow is driven by an external force $\bm{f}$, acting on forcing wavenumber $k_f$, defined in Fourier space as \mbox{$\hat{\bm{f}}(\bm{k}) = \varepsilon \hat{\bm{u}}(\bm{k}) / \sum_{k_f \leq |\bm{k}| < k_f+1} |\hat{\bm{u}}(\bm{k})|^2$}, ensuring smoothness and a constant energy injection rate $\varepsilon$ \cite{Lamorgese2005}.

Furthermore, $\bm{f}_p$ denotes the back-reaction from the point particles onto the fluid phase, which is governed by momentum exchange as \cite{Maxey1994,Mazzitelli2003,DeWit2024}
\begin{equation} \label{eq:bubble_force}
    \bm{f}_p \equiv \sum_i^{N_p} \left[ \frac{\mathrm{D} \bm{u}}{\mathrm{D} t} - \frac{\rho_p}{\rho_f} \frac{\mathrm{d} \bm{v}_i}{\mathrm{d} t} \right] \mathcal{V}_p \delta\left(\bm{r} - \bm{x}_i(t) \right),
\end{equation}
summing the individual contributions to the feedback force for all $N_p$ particles. Here, $\bm{v}_i$ and $\bm{x}_i$ denote the velocity and position of particle $i$, respectively, $\mathcal{V}_p$ is the volume of the modeled particle, and $\rho_p$ and $\rho_f$ represent the density of the particle and the fluid, respectively. For massless bubbles, $\rho_p=0$, such that only the first term in the square brackets remains, originating from the added mass of the bubble. Under the point-particle assumption, the back-reaction force is $\delta$-distributed in space, which is approximated in practice by extrapolating to the nearest eight grid points surrounding the particle, see Ref.~\cite{DeWit2024}.

Finally, the particle phase is evolved through Lagrangian interpolation following the Maxey-Riley equations \cite{Maxey1983}, where we retain only the dominant forces, being the pressure gradient force, added mass and Stokes drag, yielding
\begin{subequations}
\begin{align}\label{eq:particle}
    \frac{\mathrm{d} \bm{v}}{\mathrm{d} t} &= \beta \frac{\mathrm{D} \bm{u}(\bm{x},t)}{\mathrm{D} t} -\frac{1}{\tau_p}(\bm{v}-\bm{u}(\bm{x},t)),\\
    \frac{\mathrm{d} \bm{x}}{\mathrm{d} t} &= \bm{v} + \textrm{(collisions)}.
\end{align}
\end{subequations}
Here, $\beta\equiv3/(1+2\rho_p / \rho_f)$ and $\tau_p$ is the reciprocal Stokes drag coefficient or particle response time, which is typically compared with the Kolmogorov time $\tau_\eta\equiv\sqrt{\nu/\varepsilon}$ of the turbulence to yield the Stokes number $\textrm{St}\equiv\tau_p/\tau_\eta$. 

We resolve collisions between particles by assuming hard spherical particles with collision diameter $D$, using the `You Only Collide Once' (YOCO) algorithm. This method performs elastic collisions on particles that are forecast to collide in the next time step and re-enforces excluded-volume for particles that are overlapping, treating the most severe overlap or collision for each particle at every time step, see Ref.~\cite{DeWit2024} for a detailed description. This method presumes that collisions happen instantaneously. In reality, there will be an elastic time scale involved in the bubble collision process. Our method thus represents the limit in which this elastic time scale of collisions is much smaller than the smallest dynamical time scale of the turbulent flow, represented by $\tau_\eta$.

This work shall focus on massless bubbles ($\rho_p/\rho_f = 0$, $\beta=3$), with a resonant Stokes number $\textrm{St}=1$, yielding the strongest effect of the characteristic preferential concentration of bubbles in high-vorticity regions \cite{Calzavarini2008}. The main simulations are carried out with a resolution of $512^3$ Eulerian grid points and Taylor-scale Reynolds number $\textrm{Re}_\lambda\approx95$, while we also perform smaller simulations at $256^3$ and $128^3$ resolution and smaller $\textrm{Re}_\lambda$ to assess $\textrm{Re}$-dependence. We ensure that the largest wavenumber resolves $k_\textrm{max}\eta\approx 3$, such that the grid spacing coincides with the Kolmogorov length $\eta\equiv\nu^{3/4}\varepsilon^{-1/4}$, yielding a well-resolved dissipative range. Bubbles with various collision diameters are considered in the range $D=(0.41-1.63)\eta$, which is sufficiently small to justify the point-particle approximation \cite{homann2010finite,calzavarini2009acceleration}. While formally, the ratio $D/\eta$ would impose the $\textrm{St}$ if one assumes pure spherical Stokesian drag, we numerically fix $\textrm{St}=1$ for all bubble sizes. This not only maximizes the effect of preferential concentration, but also isolates the effect of bubble-bubble interactions from inertial effects, while ensuring that bubbles can be modeled to be sufficiently small. To assess the influence of bubble volume fraction, the number of particles is varied over a large dynamic range from $N_p=256$ to $N_p=4\times10^7$, yielding a bubble volume fraction $\alpha\equiv \pi N_p D^3 / (6 \mathcal{V})$ (with $\mathcal{V}$ the total simulation volume) that varies all the way from a very dilute regime with $\alpha=\mathcal{O}(10^{-5})$ to a very dense regime with $\alpha=\mathcal{O}(10^{-1})$. This allows us to systematically explore the effect of four-way coupling on bubbly turbulence.

\section{Bubble-fluid interaction} \label{sec:pf} 

\begin{figure}[t!]
\includegraphics[width=0.95\textwidth]{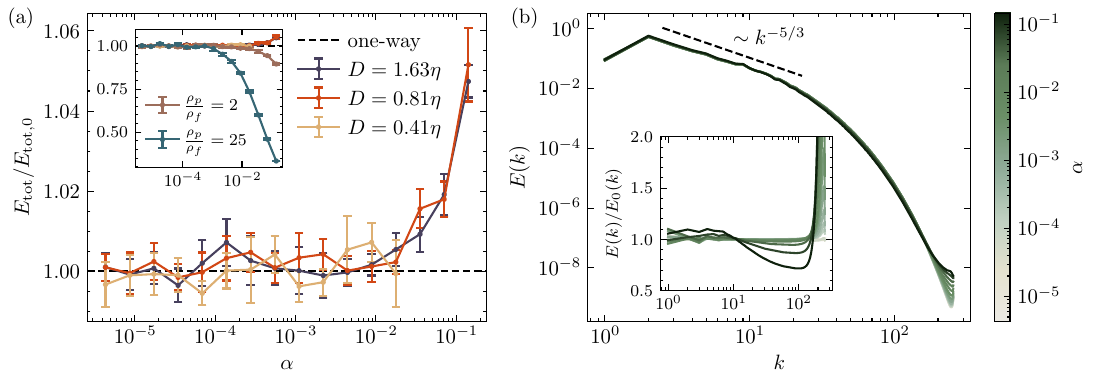}
\caption{\label{fig:kin_energy}Total kinetic energy (a) and kinetic energy spectra (b) for four-way coupled bubbles in turbulence at various volume fractions $\alpha$ and bubble sizes $D$. In (a) the kinetic energy $E_\textrm{tot}$ is compensated by the kinetic energy of the single-phase turbulence without bubbles $E_\textrm{tot,0}$, while (b) shows the total energy spectra with the compensated spectra in the inset. The inset in (a) shows the kinetic energy for runs with heavy particles ($\rho_p/\rho_f=2$ and $\rho_f/\rho_p=25$) for comparison. Panel (b) provides the results for bubbles with $D=0.81\eta$. The figure shows that the back-reaction of the bubbles on the fluid flow only has a very small energetic effect, even at relatively large volume fractions.}
\end{figure}

To understand the effect of the four-way coupled bubbles on the underlying turbulent flow, we start by studying how they affect the kinetic energy. Changes in the energetic properties of the flow will be a consequence of the back-reaction that the bubbles exert on the fluid phase. In Fig.~\ref{fig:kin_energy}, we show the total kinetic energy $E_\textrm{tot}$ (panel a) in the fluid phase for the four-way coupled system at varying bubble volume fraction, compared with its one-way coupled counterpart (with the same $\beta=3$ and $\textrm{St}=1$), as well as the corresponding spectral distribution of energy $E(k)$ as a function of wavenumber $k$ (panel b).

It shows that only when one goes to extremely high volume fraction of $\alpha \gtrsim10\%$, an appreciable effect on the total kinetic energy is obtained. However, even at these very high volume fractions, the quantitative effect on total kinetic energy remains very small, at the sub-5\% level. We furthermore point out that the effect on the kinetic energy collapses for different $D$, which is a consequence of the proportionality of the feedback force of each bubble with the bubble volume $\mathcal{V}_p$, making the total feedback force effectively proportional to the local volume fraction, see Eq.~\eqref{eq:bubble_force}. Considering the different spectral scales, we also find just small effects, which only become clear at the highest volume fractions. We notice an increase of energy at the very smallest scales (highest $k$), corresponding to the scale of the bubbles, followed by a slight attenuation of spectral energy in the dissipative range and an even slighter increase of energy in the forcing and inertial scales. This is in agreement with earlier results for two-way coupled bubbles \cite{mazzitelli2003effect}, and can be understood from basic arguments \cite{DeWit2024}: if one approximates the feedback force as $\bm{f}_p\approx \alpha \frac{\mathrm{D}\bm{u}}{\mathrm{D}t}$, see Eq.~\eqref{eq:bubble_force}, then a substitution into the Navier-Stokes Eq.~\eqref{eq:NS}, reveals that both the energy injection rate $\varepsilon$ as well as the viscosity $\nu$ are effectively enhanced by a factor $1/(1-\alpha)$, which slightly increases the energy in the inertial range and moves the Kolmogorov scale to lower $k$, thereby decreasing the energetic content in the dissipative range. In this sense, the modulation of the fluid flow due to the presence of bubbles effectively boils down to a decrease of the local fluid density by a factor ($1-\alpha$).

The relatively small magnitude of the effect of the four-way coupled bubbles on the fluid kinetic energy can be understood from the nature of the back-reaction force itself. For fluids laden with heavy particles, strong turbulence attenuation is obtained even at only moderate volume fractions \cite{boivin1998direct,ferrante2003physical,shen2022turbulence,brandt2022particle}, see also the inset in panel (a). The massless bubbles, on the contrary, do not carry any momentum themselves. As a consequence, the only back-reaction in the case of bubbles is due to their added mass, i.e. the first term in the square brackets in Eq.~\eqref{eq:bubble_force}, while the second term vanishes, yielding only a relatively small force on the underlying flow.

\section{Preferential sampling and clustering} \label{sec:pref} 

Although the back-reaction on the fluid may be small, we can expect a significant difference between four-way coupled and one-way coupled bubbles in the Lagrangian statistics due to the interactions between bubbles themselves. Because of the preferential concentration inside high-vorticity regions, local densities can be much larger than the mean density, which raises the importance of bubble-bubble interactions even further. To investigate the effect of these interactions on the preferential concentration of bubbles inside the high-vorticity regions, we study the enstrophy $\Omega^2\equiv|\bm{\nabla}\times\bm{u}|^2$ that is sampled at the position of the bubble.

\begin{figure}[t!]
\includegraphics[width=\textwidth]{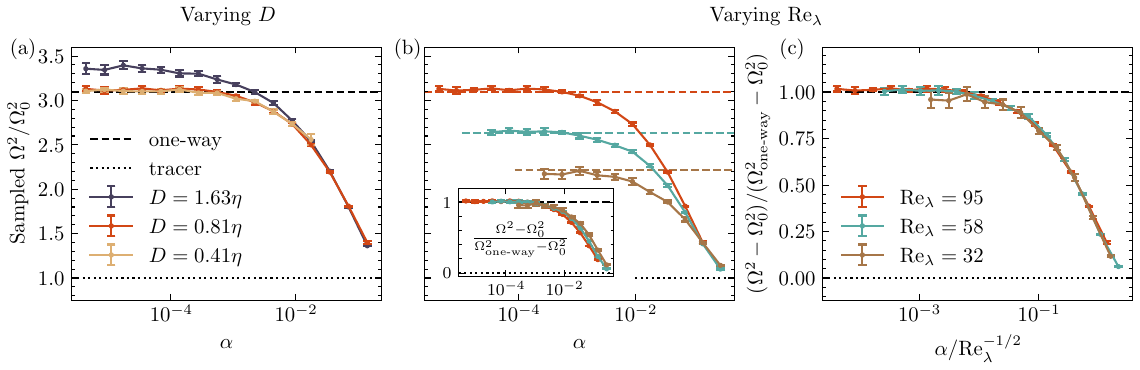}
\caption{\label{fig:sampled_ens}The effect of four-way coupling on the preferential sampling of enstrophy as a function of the volume fraction $\alpha$. The sampled enstrophy $\Omega^2$ at the bubble position is compensated by the Eulerian domain average $\Omega_0^2$, shown for varying bubble sizes $D$ (with $\textrm{Re}_\lambda=95$) (a) and for varying Taylor-scale Reynolds numbers $\textrm{Re}_\lambda$ (with $D=0.81\eta$) (b). The inset in (b) shows the preferential sampling of enstrophy rescaled with respect to the one-way coupled sampled enstrophy as $(\Omega^2-\Omega_0^2)/(\Omega_{\textrm{one-way}}^2-\Omega_0^2)$. Panel (c) shows the same data, where the volume fraction is rescaled by a factor $\textrm{Re}_\lambda^{-1/2}$. Horizontal dashed lines indicate the results for one-way coupled bubbles. The figure shows that the vortex filaments with the highest enstrophy become filled up at around a critical volume fraction of $\alpha=\alpha_c\approx 1\%$, beyond which preferential sampling is inhibited.}
\end{figure}

The results are shown in Fig.~\ref{fig:sampled_ens} for varying volume fraction $\alpha$ and for different bubble sizes $D$ and turbulence Taylor-scale Reynolds number $\textrm{Re}_\lambda$. The figure shows a clear transition between a dilute regime at low $\alpha$ and a denser regime at higher $\alpha$. In the dilute regime, the preferential sampling of the four-way coupled bubbles is indistinguishable from that of the one-way coupled bubbles. In the denser regime, preferential sampling decreases as bubbles start to interact through their excluded-volume and eventually approaches the Eulerian average expected for uniformly distributed bubbles at the highest volume fractions. The transition between the two regimes occurs at around a critical volume fraction $\alpha=\alpha_c\approx1\%$, independent of the bubble size $D$ in the considered range, as indicated by the collapse in panel (a). We recall that this collapse is with respect to the volume fraction $\alpha$, which for each bubble size corresponds to a different number of bubbles. The largest bubbles ($D=1.63\eta$) show a small excess oversampling of enstrophy beyond the one-way coupled bubbles at low $\alpha$. This is attributed to the self-induced vorticity sensed by the bubbles that results from the perturbations produced by the localized back-reaction $\bm{f}_p$ on the fluid.

The high-enstrophy regions that the bubbles preferentially sample are associated with vortex filaments in the turbulent flow \cite{douady1991,jimenez1993structure,yeung2015extreme,Calzavarini2008,Toschi2009,Aliseda2011,Wang2024}. These coherent structures are long-lived small-scale structures that extend in a line-like filamentary fashion. They have been an active topic in fundamental turbulence research and are presumed to be intricately connected to turbulence intermittency \cite{Frisch_1995,buzzicotti2020statistical,freitas2025statistical}. The findings in this work shed some light on the space occupied by these high-vorticity regions. Since we find that preferential sampling decreases from around $\alpha_c\approx1\%$ for the considered $\textrm{Re}_\lambda=\mathcal{O}(10^2)$, we can regard this as the point at which the vortex filaments become essentially `filled-up', such that less preferential locations need to be sampled, making this transition point a proxy for the volume fraction occupied by vortex filaments.

To assess the $\textrm{Re}$-dependence of this transition point, we consider the preferential sampling for different $\textrm{Re}_\lambda$ in panel (b). This yields different levels of preferential sampling due to their different levels of intermittency, but we can linearly rescale the enstrophy to be between 0 and 1 (inset). In panel (c) we show that these curves then collapse well upon rescaling the volume fraction by a factor $\textrm{Re}_\lambda^{-1/2}$. This suggests that the transition point, associated with the volume fraction of vortex filaments, scales as $\alpha_c\propto \textrm{Re}_\lambda^{-1/2}$, although it must be noted that the accessible dynamic range of $\textrm{Re}_\lambda$ is limited. We point out that this scaling relation is an empirical finding here, and how to rationalize it through scaling arguments remains an interesting open question. This would relate to both the size as well as the effective number density of vortex filaments, and would possibly also be affected by the lifetime of vortex filaments. The approach demonstrated in this work using four-way coupled bubbles seems to provide a promising pathway to study such questions. 

\begin{figure}[t!]
\includegraphics[width=0.57\textwidth]{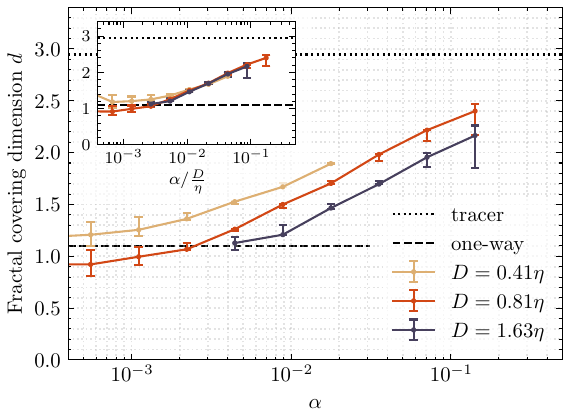}
\caption{\label{fig:fractal_dim}Fractal covering dimension $d$ for four-way coupled bubbles in turbulence at varying volume fraction $\alpha$ for different bubble sizes $D$. It shows that, as volume fraction increases, the bubble clusters transition from the filamentary line-like structures with $d\approx 1$ that are retrieved for one-way coupled bubbles to more volumetric structures with $d\to 3$. The inset shows a rescaling of the abscissa by the bubble size $D$.}
\end{figure}

As we approach this transition point at which the most preferential regions become filled, we should also expect the morphology of the bubble clusters to change. To investigate this, we compute the fractal covering dimension $d$ of the total set of bubbles, capturing the spatial dimensionality of the bubble configuration. Details on this computation are provided in Appendix~\ref{app:frac_dim}. The resulting fractal dimensions are shown in Fig.~\ref{fig:fractal_dim}. This indeed shows how the bubble cluster conformation changes from one dimensional ($d\approx 1$) line-like structures at low $\alpha$, following the morphology of vortex filaments, to a gradually more volumetric conformation at higher $\alpha$, approaching full three dimensionality ($d\to 3$). This change can also be appreciated visually from the example snapshot in Fig.~\ref{fig:4way_example}. For the fractal covering dimension, there is an explicit dependence on $D$, as larger bubbles remain in a line-like configuration until higher $\alpha$ than smaller bubbles when filling up the vortex filaments. This is also evident from the apparent empirical collapse when rescaling the abscissa by the bubble size $D$ (inset). The fractal dimension going slightly below the one-way coupling line for low $\alpha$ is attributed to the finite number of bubbles, leading to point-like contributions from individual bubbles at the lowest $\alpha$, as motivated in Appendix~\ref{app:frac_dim}.

\section{Lagrangian velocity structure function} \label{sec:lvsf} 
Another commonly studied property of Lagrangian turbulence where we may expect signatures of bubble-bubble interactions is the velocity structure function
\begin{equation}
    S_p(\tau)\equiv\langle \left| \delta_\tau\bm{v} \right|^p \rangle = \langle \left| \bm{v}(t+\tau) - \bm{v}(t) \right|^p \rangle,
\end{equation}
and its scaling behavior $ S_p(\tau)\propto\tau^{\zeta_p}$ with scaling exponents $\zeta_p$. These structure functions can be used to describe the intermittency properties of Lagrangian turbulence through the deviation from the Kolmogorov prediction $\zeta_p = p/2$ \cite{Kolmogorov1941,kraichnan1965lagrangian}.

\begin{figure}[t!]
\includegraphics[width=0.6\textwidth]{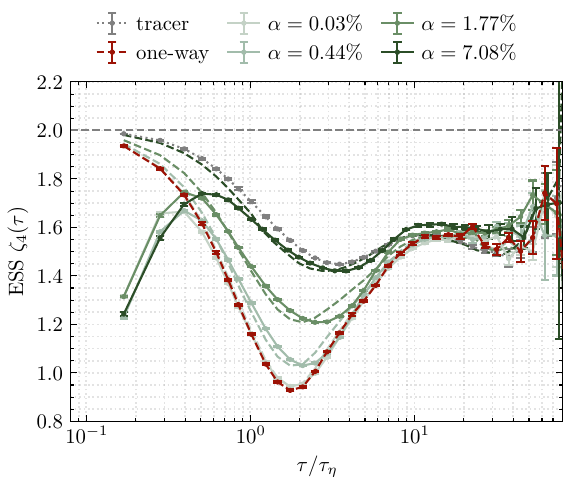}
\caption{\label{fig:lvsf}Lagrangian velocity structure function scaling exponent $\zeta_4(\tau)$ for four-way coupled bubbles in turbulence with $D=0.81\eta$ at various volume fractions $\alpha$. The exponents are obtained via extended self-similarity (ESS) as $\zeta_4(\tau) \equiv \mathrm{d}\log S_4 (\tau)/\mathrm{d}\log S_2 (\tau)$. Green dashed lines indicate fits through linear combination $\zeta_4^\textrm{(fit)} = \gamma \zeta_4^\textrm{(tracer)} + (1-\gamma)\zeta_4^\textrm{(one-way)}$. The horizontal dashed line indicates the Kolmogorov non-intermittent smooth solution $\zeta_4=2$. It shows that the viscous dip around $\tau = \mathcal{O}(\tau_\eta)$, that is amplified for bubbles due to preferential sampling, is reduced as the volume fraction increases.}
\end{figure}

To measure the scaling exponents $\zeta_p$, they are typically obtained using extended self-similarity (ESS) \cite{benzi1993extended,mordant2001measurement,biferale2004multifractal} as
\begin{equation}
    \zeta_p(\tau) \equiv \frac{\mathrm{d} \log S_p(\tau)}{\mathrm{d} \log S_2(\tau)}.
\end{equation}
The resulting local scaling exponents $\zeta_p(\tau)$ of the Lagrangian velocity structure functions are known to show a notable decrease around the dissipative scale $\tau = \mathcal{O}(\tau_\eta)$, before plateauing into the inertial range at $\tau>\mathcal{O}(10\tau_\eta)$ \cite{biferale2008lagrangian}. This so-called `viscous dip' is associated with trapping in vortex filaments \cite{biferale2005particle,bec2006effects} and is hence amplified for light particles \cite{mazzitelli2004lagrangian}.

In Fig.~\ref{fig:lvsf}, we show the obtained ESS Lagrangian velocity structure function scaling exponents of the 4th order for the four-way coupled bubbles, alongside the one-way coupled bubbles and tracer particles for comparison. It shows that at low volume fraction $\alpha \lesssim 1 \%$, the scaling exponents closely follow that of the one-way coupled bubbles, while at larger volume fractions $\alpha \gtrsim 1 \%$ the scaling exponents get closer to that of tracers. At these higher volume fractions, preferential sampling and thus trapping in vortex filaments decreases, resulting in a less pronounced viscous dip. At larger time scales, approaching the inertial range $\tau \gtrsim 10 \tau_\eta$, all scaling exponents collapse within error bars. On the other hand, at the shortest time scales $\tau \ll \tau_\eta$, while one-way coupled bubbles and tracers recover the smooth solution $\lim_{\tau \to 0}\zeta_4(\tau) \approx 2$, the four-way coupled bubbles approach the rough, shock-like solution $\lim_{\tau \to 0} \zeta_4 (\tau) \approx 1$ due to the effect of the instantaneous collisions, akin to Burgers turbulence \cite{burgers1948,de2023dynamic}. Since in this short time limit, no other dynamical processes are active, these collisions quickly dominate the statistics of velocity differences at the shortest time scales even for the smallest volume fractions when collisions are increasingly rare. However, crucially, as can be seen from the figure, the effect of collisions quickly vanishes as $\tau$ approaches $\tau_\eta$ and beyond. This indicates that results obtained for one-way coupled bubbles are correctly approaching those for four-way coupled bubbles in the dilute limit $\alpha \lesssim 1 \%$ at all relevant dynamical time scales when it comes to the Lagrangian velocity structure function. In the non-dilute regime $\alpha \gtrsim 1\%$, the scaling exponents are well-described by a simple linear combination of the scaling exponents of tracers and one-way coupled bubbles $\zeta_4^\textrm{(fit)} = \gamma \zeta_4^\textrm{(tracer)} + (1-\gamma)\zeta_4^\textrm{(one-way)}$, with fit parameter $\gamma$, as indicated by the green dashed lines in Fig.~\ref{fig:lvsf}.

\section{Pair dispersion} \label{sec:pair} 
\begin{figure}[t!]
\includegraphics[width=0.9\textwidth]{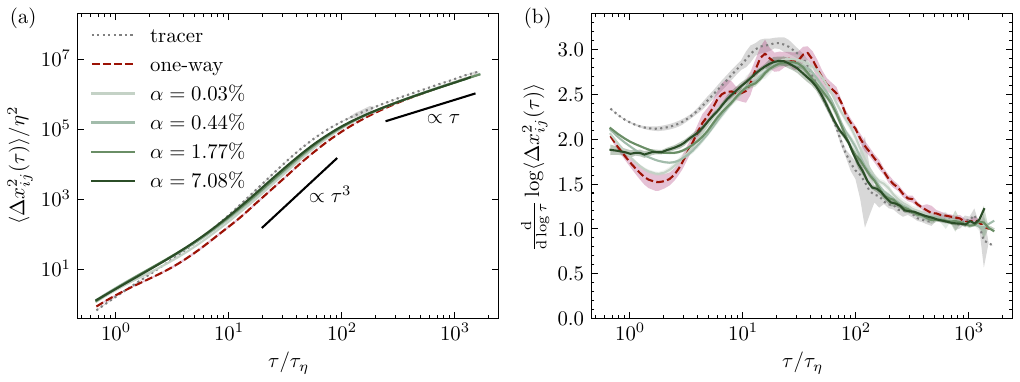}
\caption{\label{fig:pair_dispersion}Pair dispersion for four-way coupled bubbles in turbulence with $D=0.81\eta$ at various volume fractions $\alpha$. The figure shows the mean squared displacement $\langle\Delta x_{ij}^2 (\tau)\rangle$ (a) and its logarithmic derivative (b) of bubble pairs with initial separation $0.9D < \Delta x_{ij}(0) < 3.5\eta$ retrieving the Richardson dispersion results with an exponent that reaches $\approx 3$ in the inertial range, before dropping to $\approx 1$ in the diffusive range. It shows that the mean pair dispersion remains largely unaffected by the four-way coupling.}
\end{figure}

Finally, we study how the four-way coupling affects the dispersion of pairs of bubbles in the turbulent flow. To that extent, we consider the evolution of the separation between pairs of bubbles $\bm{x}_i$ and $\bm{x}_j$ as
\begin{equation}
    x_{ij} (\tau) \equiv | \bm{x}_i(t+\tau) - \bm{x}_j(t+\tau) |,
\end{equation}
which initially are close together $0.9D<x_{ij}(0) < 3.5\eta$. To eliminate the effect of the initial condition, we study the quantity
\begin{equation}
    \Delta x_{ij}(\tau) \equiv x_{ij}(\tau) - x_{ij}(0).
\end{equation}
In Fig.~\ref{fig:pair_dispersion}, we plot the mean squared displacement $\langle\Delta x_{ij}^2 (\tau)\rangle$ and its scaling behavior for four-way coupled bubbles at different volume fractions as well as for tracers and one-way coupled bubbles for comparison. The pair dispersion shows an initial ballistic growth, followed by the famous Richardson superdiffusive growth as $\propto\tau^3$ in the inertial range, and a diffusive regime $\propto\tau$ at long time beyond the integral scale ($\approx 10^2\tau_\eta$) \cite{richardson1926atmospheric,salazar2009two,bec2010turbulent}. Note that the tracers, one-way, and four-way coupled bubbles exhibit similar mean dispersion. Apart from small differences at short time scales $\sim \mathcal{O}(\tau_\eta)$ the one-way and four-way coupled bubbles largely collapse within the statistical accuracy. This indicates that the collisions between bubbles do not strongly alter their mean dispersive behavior.

\begin{figure}[t!]
\includegraphics[width=0.9\textwidth]{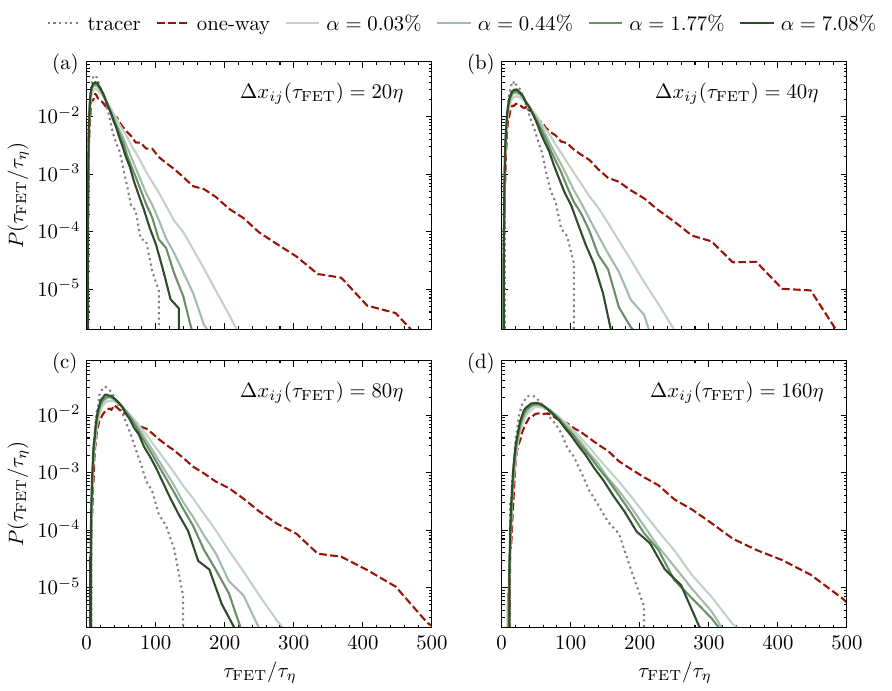}
\caption{\label{fig:first_passage_time}Distribution of first-exit times $\tau_\textrm{FET}$ for pair dispersion of four-way coupled bubbles in turbulence with $D=0.81\eta$ at various volume fractions $\alpha$ for different separation distances $\Delta x_{ij}$, respectively $20\eta$ (a), $40\eta$ (b), $80\eta$ (c) and $160\eta$ (d). It shows that, although mean pair dispersion is largely unaffected by four-way coupling, there is a strong effect on the exponential tails of the first-exit times, that capture the rare events of pairs that remain close together for an exponentially long time. Excluded-volume interactions in four-way coupling break up such long-lived pairs much more quickly than in the one-way coupling paradigm.}
\end{figure}

Although the mean dispersive behavior is not affected by bubble collisions, we may expect differences in the distribution of the pair dispersion, particularly at short distances when collisions should be more prevalent. To study this in more detail, we consider the distribution of first-exit time $\tau_\textrm{FET}$, defined as the first time when a pair of bubbles reaches a certain separation $\Delta x_{ij}$ \cite{artale1997dispersion,biferale2005lagrangian}. The resulting distributions for various separations are shown in Fig.~\ref{fig:first_passage_time}.

This shows that, although the mean dispersive behavior is not strongly affected by the four-way coupling, the tails of the distribution of the first-exit times are becoming strongly reduced. These exponential tails correspond to the rare events of pairs of bubbles that remain together for an exponentially long time, before separating and crossing the first-exit threshold. The occurrence of such long-lived pairs is strongly reduced due to the excluded-volume interactions as bubbles push each other away, even at the smallest volume fractions, explaining the large differences in the tails of the distributions. These differences gradually diminish at larger separation distances as collisions play a smaller role, as shown in the figure.

Although the contributions of these rare events of long-lived pairs to the mean dispersive behavior is negligibly small, as shown above, the presence or absence of such long-lived pairs may still have practical implications e.g. for reactive flows and four-way coupling can thus be an important consideration in the modeling for such applications at all volume fractions.

\section{Conclusions and outlook} \label{sec:conclusion} 
We have studied how the dynamics of small bubbles in turbulence changes when one goes from the commonly used one-way coupling paradigm to a four-way coupled method, where both the back-reaction from the bubbles onto the underlying fluid is considered, as well as excluded-volume interactions between the bubbles themselves. We have found that the energetic properties of the underlying turbulent flow remain largely unaffected by the presence of small passive bubbles. This can be understood from the fact that massless bubbles carry no momentum, and the back-reaction thus consists only of the added mass effect, in contrast with dispersed heavy particles. Considering the Lagrangian dynamics of the bubbles themselves, the picture becomes more subtle, as some Lagrangian statistics are found to be sensitive to the collisions between bubbles. We find a cross-over between a dilute regime at low volume fractions $\alpha \lesssim 1\%$, where collisions generally play only a small role, and a denser regime at $\alpha \gtrsim 1\%$, where the preferential concentration in filamentary high-vorticity regions reduces and more volumetric bubble clusters are retrieved. This also teaches us something about the turbulent properties of the underlying flow, suggesting that at this transition point of $\alpha = \alpha_c \approx 1\%$ for the considered $\textrm{Re}_\lambda=\mathcal{O}(10^2)$, the preferential vortex filament regions start to fill up, making it a proxy for the volume fraction occupied by such coherent vortex structures. This holds potential for applications in probing turbulence in experimental settings where one would like to measure this volume fraction in different types of turbulent flows. We find that this critical volume fraction has a weak $\textrm{Re}$-dependence, observing a scaling of $\alpha_c\propto\textrm{Re}_\lambda^{-1/2}$.

Furthermore, we show that the excluded-volume interactions can alter the Lagrangian velocity structure function. Firstly, collisions introduce roughness at the shortest sub-Kolmogorov time scales, which eventually dominates the statistics even at the smallest volume fractions as there is no other dynamics at play at such small time scales. In the dissipative and inertial range, for the dilute regime, the statistics largely follow that of the one-way coupled simulations, reassuring us of the validity of previously obtained results under the one-way coupling paradigm. As the volume fraction increases, however, the statistics gradually get closer to that of tracer particles as preferential concentration decreases, reducing the well-known `viscous dip' of the scaling exponent. Finally, considering the pair dispersion of bubbles in turbulence, we find that the mean dispersion is largely unaltered by the introduction of four-way coupling, again supporting the validity of one-way coupling simulations when studying the dispersive properties of bubbly turbulence. However, a clear signature of bubble collisions can still be observed from the tails of the distribution of first-exit times, showing that the prevalence of rare events of exponentially long-lived pairs is reduced by bubble collisions, even at the smallest volume fractions. This can have consequences for certain applications that are sensitive to such long-lived pairs, making four-way coupling an important consideration for such applications at all volume fractions.

Our findings show that the question whether or not four-way coupling is important in bubbly turbulence is a subtle question to answer and is intrinsically dependent on the type of observable that one is interested in. While we have shown that some observables are virtually insensitive to four-way coupling at all volume fractions, other observables show a dilute, unaffected regime and a denser, affected regime, while still other observables show effects of four-way coupling at all considered volume fractions. Note, however, that we have here considered massless bubbles with $\textrm{St}=1$, which show the strongest effect of preferential concentration. Moving to different $\textrm{St}$, higher or lower, will naturally result in a diminishing effect of preferential concentration and hence also the importance of the effects of four-way coupling observed here are expected to decrease as local densities decrease. A rigorous study of how preferential concentration depends on $\textrm{St}$ and density contrast $\beta$ is provided in Ref.~\cite{Calzavarini2008}.

We emphasize that this study is focusing on the effect of four-way coupling on passive bubbles that are small with respect to the Kolmogorov scale of the flow and we are not considering the effects of deformation, break-up and coalescence, nor the effect of gravity. Significant two-way interactions between the fluid and bubble phase that measurably alter the turbulent flow are expected when one considers larger, fully interacting bubbles that extend into the inertial range of the turbulence and/or are significantly affected by buoyancy \cite{prakash2012gravity,mathai2020bubbly,xu2024intermittency,ni2024deformation}. However, even in the idealized setting of small rigid bubbles studied here, a significant, well-targeted alteration of the statistics of the underlying turbulent flow can be achieved as soon as one considers external forcing on the bubbles, e.g. due to inertial or acoustic forces, as shown in Ref.~\cite{freitas2025statistical}. Future work will further explore the possibilities of this approach of targeted and controlled shaping of turbulence.

\begin{acknowledgments}
We thank Luca Biferale for useful discussions. This work is supported by the Netherlands Organization for Scientific Research (NWO) through the use of supercomputer facilities (Snellius) under Grant No. 2023.026, 2025.008. This publication is part of the project “Shaping turbulence with smart particles” with Project No. OCENW.GROOT.2019.031 of the research program Open Competitie ENW XL which is (partly) financed by the Dutch Research Council (NWO). A.F. acknowledges support from the European Union’s HORIZON MSCA Doctoral Networks programme under Grant Agreement No. 101072344, project AQTIVATE (Advanced computing, QuanTum algorIthms and data-driVen Approaches for science, Technology and Engineering) and the European Research Council (ERC) under the European Union’s Horizon 2020 research and innovation programme Smart-TURB (Grant Agreement No. 882340).
\end{acknowledgments}

\appendix

\section{Computation of fractal covering dimension}\label{app:frac_dim}

The fractal covering dimension of a set of points can be computed by tiling the embedding space with boxes of varying size $l$ and computing the number of non-empty boxes $n(l)$ that have at least one point inside. The fractal covering dimension $d$ is then readily obtained from the logarithmic derivative of $n(l)$ as
\begin{equation}
    d = -\frac{\mathrm{d} \log n(l)}{\mathrm{d}\log l}.
\end{equation}
This shows that the fractal covering dimension $d$ is in principle a function of $l$, and we are typically interested in the limiting behavior of the fractal dimension for $l\to 0$. In practice, however, this poses problem whenever considering a finite set of sample points, because at low $l$, zero-dimensional contributions from individual points can start to dominate the statistics of $n(l)$ due to the finiteness of the sample, leading to an underestimation of the true fractal dimension of a hypothetically infinite sample.

\begin{figure}[b!]
\includegraphics[width=0.99\textwidth]{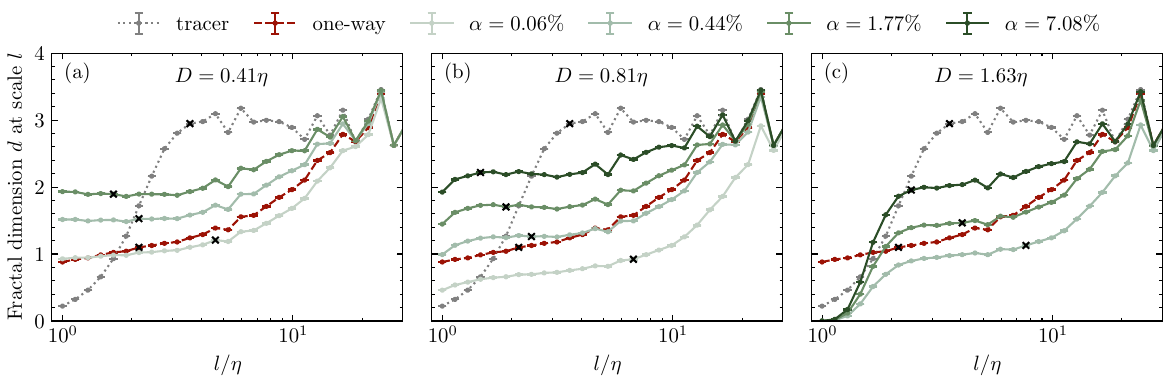}
\caption{\label{fig:frac_dim_SI} Fractal covering dimension $d$ computed at different length scales $l$ for bubbles with $D=0.41\eta$ (a), $D=0.81\eta$ (b) and $D=1.63\eta$ (c). Fractal dimensions below unity can be attributed to contributions from point-like structures resulting from the finite number of particles. The black crosses denote the selected representative fractal dimension used for the results in Fig.~\ref{fig:fractal_dim}.}
\end{figure}

This is illustrated from our data in Fig.~\ref{fig:frac_dim_SI}, showing that at low $l$, and particularly at lower $\alpha$ and larger $D$ (fewer bubbles), the fractal dimension goes below unity due to contributions from individual points. The true fractal dimension is arguably retrieved in the plateau at moderate $l/\eta$. To account for the effect of the finiteness of the number of bubbles, we select the true representative fractal dimension (used in Fig.~\ref{fig:fractal_dim}) at a length $l$ that scales as the average interparticle distance $l \propto \alpha^{-1/3}$, indicated by the black crosses in Fig.~\ref{fig:frac_dim_SI}. We propagate the uncertainty due to this scale selection into the error bar in Fig.~\ref{fig:fractal_dim}.

For the smallest $D=0.41\eta$, the highest $\alpha$ values are not accessible because of computational limitations due to the extremely high number of particles, while for the largest $D=1.63\eta$, the lowest $\alpha$ values are not reliably measurable because of the large interparticle distance due to the low number of particles.

\bibliography{main}

\end{document}